\documentclass[useAMS,usenatbib,referee]{biom}

\usepackage[figuresright]{rotating}
\usepackage{moreverb}
\usepackage{enumerate}
\usepackage{adjustbox}

\usepackage{amsfonts}
\usepackage{amsbsy,amsmath}
\usepackage{url}
\usepackage{hyperref}
\usepackage{ulem}
\usepackage{bm}
\usepackage{mathrsfs}
\usepackage{booktabs,threeparttable}
\usepackage{footnote}
\usepackage{graphicx}
\usepackage[table]{xcolor}
\usepackage{amsmath}
\DeclareMathOperator*{\argmax}{arg\,max}

\def\bSig\mathbf{\Sigma}

\usepackage[figuresright]{rotating}
\usepackage{moreverb}
\usepackage{enumerate}
\usepackage{adjustbox}
\usepackage{amsfonts}
\usepackage{amsbsy,amsmath}
\usepackage{url}
\usepackage{hyperref}
\usepackage{ulem}
\usepackage{bm}
\usepackage{mathrsfs}
\usepackage{booktabs,threeparttable}
\usepackage{footnote}
\usepackage{graphicx}
\usepackage[table]{xcolor}
\usepackage{amsmath}

\def\bSig\mathbf{\Sigma}

\newcommand{\R}{{\mathcal R}}

\def\star{^{\mbox{\tiny{\sf  *}}}}

\def\R{\mathbb R}

\def\bfx{{\ensuremath{\bf x}}}

\def\bfy{{\ensuremath{\bf y}}}

\def\bfs{{\ensuremath{\bf s}}}
\def\bfg{{\ensuremath{\bf g}}}
\def\bfA{{\ensuremath{\bf A}}}

\def\bfC{{\ensuremath{\bf C}}}
\def\bfD{{\ensuremath{\bf D}}}
\def\bfG{{\ensuremath{\bf G}}}

\def\bfN{{\ensuremath{\bf N}}}

\def\bfQ{{\ensuremath{\bf Q}}}
\def\bfS{{\ensuremath{\bf S}}}
\def\bfU{{\ensuremath{\bf U}}}
\def\bfV{{\ensuremath{\bf V}}}
\def\bfW{{\ensuremath{\bf W}}}

\def\bfX{{\ensuremath{\bf X}}}
\def\bfY{{\ensuremath{\bf Y}}}

\def\bftheta{{\ensuremath\boldsymbol{\theta}}}

\def\bfzeta{{\ensuremath\boldsymbol{\zeta}}}

\def\bfbeta{{\ensuremath\boldsymbol{\beta}}}
\def\bfgamma{{\ensuremath\boldsymbol{\gamma}}}
\def\bfomega{{\ensuremath\boldsymbol{\omega}}}

\def\bfOmega{{\ensuremath\boldsymbol{\Omega}}}

\def\bfSigma{{\ensuremath\boldsymbol{\Sigma}}}

\def\bfmu{{\ensuremath{{\boldsymbol{\mu}}}}}

\def\bfx{{\ensuremath{\bf x}}}

\def\bfy{{\ensuremath{\bf y}}}

\def\bfs{{\ensuremath{\bf s}}}
\def\bfg{{\ensuremath{\bf g}}}
\def\bfA{{\ensuremath{\bf A}}}

\def\bfG{{\ensuremath{\bf G}}}

\def\bfQ{{\ensuremath{\bf Q}}}
\def\bfS{{\ensuremath{\bf S}}}

\def\bfU{{\ensuremath{\bf U}}}
\def\bfV{{\ensuremath{\bf V}}}
\def\bfW{{\ensuremath{\bf W}}}
\def\bfE{{\ensuremath{\bf E}}}

\def\bfJ{{\ensuremath{\bf J}}}
\def\bfX{{\ensuremath{\bf X}}}
\def\bfY{{\ensuremath{\bf Y}}}

\def\bfI{{\ensuremath{\bf I}}}
\def\bfH{{\ensuremath{\bf H}}}

\def\bfbeta{{\ensuremath\boldsymbol{\beta}}}
\def\bfgamma{{\ensuremath\boldsymbol{\gamma}}}

\def\bfmu{{\ensuremath\boldsymbol{\mu}}}

\def\bftheta{{\ensuremath\boldsymbol{\theta}}}
\def\bfkappa{{\ensuremath\boldsymbol{\kappa}}}

\def\bfOmega{{\ensuremath\boldsymbol{\Omega}}}

\def\bfSigma{{\ensuremath\boldsymbol{\Sigma}}}

\def\bfepsilon{{\ensuremath\boldsymbol{\epsilon}}}

\def\bfPhi{{\ensuremath\boldsymbol{\Phi}}}
\def\bfrho{{\ensuremath\boldsymbol{\rho}}}

\title[Chen et al.]{Empirical-likelihood-based criteria for model selection on marginal analysis of longitudinal data with dropout missingness}

\author{Chixiang Chen$^{1}$, Biyi Shen$^{1}$, Lijun Zhang$^{2}$, Yuan Xue$^{3}$, Ming Wang$^{*1}$\\
	$^{1}$Division of Biostatistics and Bioinformatics, Department of Public Health Sciences \\ Penn State College of Medicine, Hershey, PA, U.S.A \\ $^{2}$Institute for Personalized Medicine, Penn State College of Medicine, Hershey, PA, U.S.A
	\\   $^{3}$ School of Statistics, University of International Business and Economics, Beijing, China\\ 
	{$^{*}$Contact Email: mwang@phs.psu.edu}}
\begin{document}

	\pubyear{2019}
	\volume{}
	\artmonth{}
	%\doi{10.1111/j.1541-0420.2005.00454.x}
	
	%  This label and the label ``lastpage'' are used by the \pagerange
	%  command above to give the page range for the article
	
	\label{firstpage}
	
	%  pub the summary here
	
	\begin{abstract}
		Longitudinal data are common in clinical trials and observational studies, {where missing outcomes due to dropouts are always encountered}. Under such context with the assumption of missing at random, {the weighted generalized estimating equations (WGEE) approach is widely adopted for marginal analysis}. {Model selection on marginal mean regression} is a crucial aspect of data analysis, and identifying an appropriate correlation structure for model fitting may also be of interest and importance. However, {the existing information criteria for model selection in WGEE have limitations}, such as separate criteria for the selection of marginal mean and correlation structures, unsatisfactory selection performance in small-sample set-ups and so on. In particular, {there are few studies to develop joint information criteria for selection of both marginal mean and correlation structures}. In this work, by embedding empirical likelihood into the WGEE framework, we propose two innovative information criteria named a joint empirical Akaike information criterion (JEAIC) and a joint empirical Bayesian information criterion (JEBIC), which can simultaneously select the variables for marginal mean regression and also correlation structure. {Through extensive simulation studies,} these empirical-likelihood-based criteria exhibit robustness, flexibility, and outperformance compared to the other criteria including the weighted quasi-likelihood under the independence model criterion, the missing longitudinal information criterion and the joint longitudinal information criterion. {In addition, we provide a theoretical justification of our proposed criteria, and present two real data examples in practice for further illustration.} 
	\end{abstract}
	
	%
	%  Please place your key words in alphabetical order, separated
	%  by semicolons, with the first letter of the first word capitalized,
	%  and a period at the end of the list.
	%
	
	\begin{keywords}
		Akaike information criterion; Bayesian information criterion; Empirical likelihood; Longitudinal data; Missing at random; Model selection; Weighted generalized estimating equation.
	\end{keywords}
	
	\maketitle
	
	\section{Introduction}
	\label{s:intro}
	Longitudinal data are common in clinical trials and observational studies. Due to the research interest in conducting inference on the population-level parameter estimates, generalized estimating equations (GEE) has been widely employed for marginal regression analysis, where the correlations among the observations within subjects are treated as nuisance parameters \citep{liang1986, ming2014}. {In longitudinal studies, missing data is typically encountered,} which poses challenges for model fitting and model selection. {There are three types of missing data: missing completely at random (MCAR), missing at random (MAR), and missing not at random (MNAR), depending on whether the factors related to missing probability are observed or not \citep{little2014}.} For instance, subjects may drop out of the study or are lost to follow-up due to several reasons such as drug resistance or side effects. Under such context, MAR is commonly and reasonably assumed for statistical inference. Literature has shown that {the estimates based on regular GEE} are biased for longitudinal data under MAR \citep{laird1988}. \cite{robins1995} first proposed the weighted GEE (WGEE) method for bias correction by incorporating an inverse probability weight matrix. Given the correctly specified model for missing data, the consistency of WGEE estimates {still holds even when the ``working" correlation structure is misspecified}. 
	
	{Model selection is a crucial aspect of longitudinal data analysis. Without a doubt, identifying the variables for the marginal mean structure is always essential. Also, an improper correlation structure may lead to loss of efficiency of parameter estimates. This problem has been exclusively investigated for complete longitudinal data; however, when the missing data exist, the efficiency improvement is still under exploration, but several works have shown that selecting a proper correlation structure for WGEE is somewhat promising and important \citep{gosho2014, gosho2016, shardell2008, preisser2002}.} {To accomplish these selection goals, development of model information criteria has gained substantial attention by researchers.} \cite{pan2001} first proposed one of the most popularly used information criteria, the quasi-likelihood under the independence model criterion (QIC), but it does not accommodate missing data. For longitudinal data with dropout missingness under MAR, \cite{shen2012} proposed two separate measures based on the quadratic loss function, the missing longitudinal information criterion (MLIC) and the MLIC for correlation (MLICC), for selection of marginal mean regression and correlation structures in WGEE, respectively. Another option for marginal model selection under this scenario is the weighted quasi-likelihood information criterion (QICW$_p$) by accommodating the weight matrix into QIC \citep{platt2013}. Later on, \cite{gosho2016} proposed QICW$_r$ by modifying the penalty term of QICW$_p$ for selection of both marginal mean and correlation structures. Most recently, \cite{shen2017} proposed the joint longitudinal information criterion (JLIC) with regards to the joint selection of marginal mean and correlation structures for longitudinal data with missing outcomes and covariates. However, {the aforementioned criteria have the following limitations: 1) ignoring missing data; 2) losing model selection power when different criteria for either marginal mean structure selection or correlation structure selection are implemented; 3) leading to unsatisfactory results in selection rates, particularly when {the sample size is small}} \citep{shen2012, shen2017, gosho2016}.
	
	On the other hand, the empirical likelihood approach by adopting a purely observation-based technique has recently gained more attention due {to the relaxing} of parametric {distributional} assumption, and literature has already shown its outperformance in regression analysis especially on confidence interval construction \citep{owen1988, qin1994, qin2009}. However, empirical-likelihood-based model selection criteria have not been widely investigated yet. \cite{eic1995} first proposed the empirical information criterion (EIC), {but pointed out that convergence to a proper solution was not reached in estimation, particularly when the number of estimating equations is larger than the number of parameters.} Later, {\cite{variyath2010} introduced adjusted empirical likelihood criteria, the empirical Akaike information criterion (EAIC) and the empirical Bayesian information criterion (EBIC), to guarantee the existence of a solution.} {However}, {the computational issue remains if the estimators have bounded support (e.g., a correlation coefficient). \cite{chen2012} applied empirical likelihood for only the correlation structure selection in GEE under complete longitudinal data and proposed to use plug-in estimators obtained from GEE; however, no theoretical justification of plug-in estimators was provided in their work}. To our knowledge, there is little work on empirical-likelihood-based model selection criteria accommodating missing data under the longitudinal framework. 
	
	In this paper, {two motivated data applications are provided.} One is a large epidemiological study, the Atherosclerosis Risk in Communities (ARIC) study. Systolic blood pressure (SBP), a crucial risk factor for cardiovascular disease (CVD), is of clinical and research interest, and characterizing its longitudinal patterns over time can help for CVD risk prediction and determine relatively more effective treatment or medication \citep{muntner2015, parati2013}. {The other one is a study of Schizophrenia disorder.} The mean level, as well as visit-to-visit variability on severity measurements, is associated with deficits in emotional processing and functional impairment \citep{simon2007, bilderbeck2016}, which could reflect drug effectiveness and indicate a strategy for prevention of disease progression. To achieve these clinical objectives, we need to identify the best fitting model among different candidates. Here, we propose two information criteria named a joint empirical Akaike information criterion (JEAIC) and a joint empirical Bayesian information criterion (JEBIC), which can simultaneously select marginal mean and correlation structures in WGEE for longitudinal data with dropout missingness under MAR. The basic strategy is that the empirical-likelihood-based criteria are first established by utilizing parameter estimates from WGEE together with the proposed empirical likelihood, and thus JEAIC and JEBIC can be constructed by incorporating extra penalty terms. {These criteria are easy to implement in statistical software}, and {potential computational issues} can be avoided because the parameter estimates are obtained directly from WGEE. Also, this work can be extended to accommodate more general missing patterns (i.e., intermittent missingness). For simplicity, we mainly focus on monotone dropout missingness here.
	
	The paper is organized as follows. In Section \ref{s:method}, we formulate the problem, introduce WGEE and the existing model selection criteria, and then provide the proposed information criteria of JEAIC and JEBIC based on the empirical likelihood. The theoretical justification {for} our proposal is granted under certain conditions with detailed proof in {the Supporting Information}. In Section \ref{s:simulation}, we conduct extensive simulations under a variety of scenarios with continuous and categorical outcomes {to evaluate the performance of the two proposed criteria when compared with the current existing alternatives}. Lastly, we illustrate the application of our scheme by utilizing two real data examples in Section \ref{s:example}, and conclude with a discussion in Section \ref{s:discussion}.
	
	\vspace{-0.3in}
	\section{Methodology}
	\label{s:method}
	\subsection{Notation}
	Let $\bfY_i = (Y_{i1},\dots, Y_{iT})^\prime$ and $\bfX_i = (\bfX_{i1},\dots, \bfX_{iT})^\prime$ {denote the outcomes and covariates} collected from subject $i, i=1, \dots, n$, respectively, {where $Y_{ij}$ is the $j^{th}$ outcome and a $p \times 1$ vector of covariates $\bfX_{ij}$ includes the intercept,} $j = 1,\dots, T$. For simplicity, we assume balanced data with equal numbers of observations for all subjects. Let $\bfmu_i = E(\bfY_i |\bfX_i)$ and $\bfV_i = Var(\bfY_i |\bfX_i)$ be {the conditional mean and variance of $\bfY_i$}. Note that $\bfmu_i$ is usually modeled {as} $\xi(\bfmu_i)=\bfX_i \bfbeta$ with $\xi$ as a known and pre-specified link function depending on the type of outcomes and $\bfbeta$ as a $p \times 1$ vector of regression parameters \citep{mccullagh1989}. In addition, $\bfV_i$ can be written by $\bfA^{1/2}_i \bfC_i(\bfrho) \bfA^{1/2}_i$, where the matrix $\bfA_i$ is a $T \times T$ diagonal matrix with diagonal elements $var(Y_{it} |\bfX_{it})= \phi \nu (\mu_{it})$, where $\nu$ is a known function, and $\phi$ is a dispersion parameter which could be known or has to be estimated if unknown; $\bfC_i (\bfrho)$ is a pre-specified ``working" correlation matrix depending on a set of parameters $\bfrho$. Here, we consider the outcomes subject to missingness under the assumption of MAR, {where the indicator $R_{ij}=1$ for the observed $Y_{ij}$ and $R_{ij}=0$, otherwise}. For simplicity, we focus on dropout missingness, but it can be straightforwardly extended to accommodate other general missing patterns \citep{robins1995, shen2017}. 
	
	\subsection{WGEE} \label{section 2.2}
	For longitudinal data with dropouts under MAR, WGEE has been proposed by incorporating a weight matrix based on the inverse probability of observing the outcomes to {adjust for the missing mechanism} \citep{robins1995}. Let the probability of observing the outcome for the $i^{th}$ subject as $\bfomega_{i}=(\omega_{i1}, \ldots, \omega_{iT})^\prime$, where $\omega_{ij}=Pr(R_{ij}=1| \bfY_i, \bfH_i)$ {with $\bfH_{i}$ including potential predictors which could be overlapped with $\bfX_i$.} Note that $\omega_{ij}=\lambda_{i1} \times \lambda_{i2}\times \dots \times \lambda_{ij}$ where $\lambda_{i1}=1$ (the outcomes at baseline are all observed) and $\lambda_{ij}=Pr(R_{ij}=1|R_{i,j-1}=1, \bfY_i, \bfH_{i}), j=2, \dots, T$. Given the data $(R_{ij}, \bfY_i, \bfH_{i})$, $\lambda_{ij}$ can be estimated based on the partial likelihood from a logistic regression, 
	$\sum_{i=1}^{n} \sum_{j=2}^{T} R_{i,j-1}log[\lambda_{ij}(\bftheta)^{R_{ij}} \{1-\lambda_{ij}(\bftheta)\}^{1-R_{ij}} ]$, where $\bftheta$ is a $q \times 1$ vector of regression parameters with consistent estimates obtained by
	\begin{equation}\label{ees}
	\bfS_{n\bftheta}=\frac{1}{n}\sum_{i=1}^{n}\bfs_i(\bftheta)=\frac{1}{n}\sum_{i=1}^{n}\sum_{j=2}^{T}R_{i,j-1}\Big\{R_{i,j}-\lambda_{ij}(\bftheta)\Big\}\bfH_{ij},
	\end{equation}
	with $\text{logit}\big(\lambda_{ij}(\bftheta)\big)=\bfH_{ij}^\prime\bftheta$. Thus, the predicted probability $\widehat{\lambda}_{ij}$ and thereafter $\widehat{\omega}_{ij}$ can be calculated. After plugging $\widehat{\bfomega}$ into $\bfW_i$, the estimating equations for the parameters $\bfbeta$ are 
	\begin{equation}\label{eebeta}
	g(\bfbeta)=\sum_{i=1}^{n} g(\bfX_i, \bfY_i, \bfbeta; \widehat{\bfomega})=\sum_{i=1}^{n}\bfD_i^{\prime} \bfV_i^{-1} \bfW_i(\bfY_i-\bfmu_i)=0,
	\end{equation}
	where $\bfD_i=\partial \bfmu_i/\partial \bfbeta^{\prime}$ which is a $T \times p$ matrix, $\bfV_i=\bfA_i^{1/2}\bfC_i \bfA_i^{1/2}$, and $\bfW_i$ is the weight matrix with diagonal elements $R_{ij}/\widehat{\omega}_{ij},j=1, \dots, T$. The estimate $\widehat{\bfbeta}$ is consistent even if the ``working" correlation matrix is misspecified, and $\sqrt{n}(\widehat{\bfbeta}-\bfbeta)$ is asymptotically normal distributed under mild regulatory conditions, given that the dropout model is correctly specified (i.e., $E \bfW_i=\bfI_T$, with $\bfW_i$ evaluated at the true value $\bfomega_0$) \citep{robins1995}. 
	
	Note that given any pre-specified ``working" correlation matrix $\bfC$ other than an independent correlation structure, the correlation coefficient $\bfrho$ needs to be estimated. Usually, the correlation estimates can be obtained based on an iterative process by utilizing the Pearson residuals \citep{wedderburn1974}. But, the correlation coefficient estimate for the longitudinal data with missing outcomes could be biased, while the unbiased estimate for $\rho_{jk}$ is 
	$\widehat{\rho}_{jk}(\widehat{\bfbeta})=[\{1/\{(n-p)\phi\}]\sum_{i=1}^{n}{e_{ij}(\widehat{\bfbeta})e_{ik}(\widehat{\bfbeta})R_{ij}R_{ik}/\widehat{\omega}_{i,jk}} $
	where $\widehat{\omega}_{i,jk}$ is the estimate of $\omega_{i,jk}=Pr(R_{ij}=1, R_{ik}=1| \bfY_i, \bfH_{ij}, \bfH_{ik})$ and $e_{ij}(\bfbeta)$ is the residual $(Y_{ij}-\mu_{ij})/\sqrt{\nu(\mu_{ij}})$ ($1\leq j<k\leq T$). Because of dropout missingness, the weights can be simplified as $\omega_{i,jk}=\omega_{ik}=Pr(R_{ik}=1| \bfY_i, \bfH_{ik})$ and then $\widehat{\rho}_{jk}(\widehat{\bfbeta})=[1/\{(n-p)\phi\}]\sum_{i=1}^{n}{e_{ij}(\widehat{\bfbeta})e_{ik}(\widehat{\bfbeta})R_{ik}/\widehat{\omega}_{ik}}$; {For other missing patterns (i.e., intermittent), the estimation would become more complicated \citep{robins1995, chen2010}.} In addition, $\phi$ is assumed to be known or estimated as $\widehat{\phi}(\widehat{\bfbeta})=\{1/(nT-p)\}\sum_{i=1}^{n}\sum_{j=1}^{T}e^2_{ij}(\widehat{\bfbeta})R_{ij}/\widehat{\omega}_{ij}$ (released afterwards for mathematical simplicity). For convenient notation, we stack the estimating equations by subject $i$ for the parameters $\bfgamma=(\bfbeta^\prime, \bfrho^\prime)^{\prime}$ as follows
	\begin{align}\label{wgee}
	\bfg\Big(\bfX_i, \bfY_i, \bfgamma; \widehat{\bfomega}_i\Big)=&
	\begin{pmatrix} 
	\bfD_i^\prime \bfV_i^{-1}\bfW_i\Big\{\bfY_i-\bfmu_i(\bfbeta)\Big\}\\  
	\bfzeta(\bfX_i,\bfY_i, \bfrho; \widehat{\bfomega}_i)
	\end{pmatrix},
	\end{align}
	where $\bfzeta(\bfX_i,\bfY_i, \bfrho; \widehat{\bfomega}_i)$ is some estimating equation for the correlation coefficients $\bfrho$ based on weighted Pearson residuals. Taking an unstructured case for example, $\bfzeta(\bfX_i,\bfY_i, \bfrho; \widehat{\bfomega}_i)$ could be $\bfkappa_i(\bfbeta)-\bfrho\phi(1-p/n)$, where $\bfkappa_i(\bfbeta)=\big(\widehat{\rho}_{i12}(\bfbeta), \ldots, \widehat{\rho}_{i1T}(\bfbeta), \ldots, \widehat{\rho}_{i(T-1)T}(\bfbeta)\big)^{\prime}$ with $\widehat{\rho}_{ijk}(\bfbeta)=e_{ij}(\bfbeta)e_{ik}(\bfbeta)R_{ik}/\widehat{\omega}_{ik}, 1\leq j<k\leq T$, and $\bfrho=\big(\rho_{12},\cdots,\rho_{1T},\cdots,\rho_{(T-1)T}\big)^\prime$. %The estimates of $\bfgamma$ can be obtained by an iterative algorithm. 
	
	\subsection{Model selection criteria}\label{section 2.3}
	\subsubsection{Overview of Existing Criteria}\label{section2.3.1}
	Before introducing our proposed information criteria, we first {conduct a literature review of several key criteria on model selection for WGEE in longitudinal data analysis, with dropout missingness under MAR.} One called MLIC was proposed for the selection on marginal mean regression by \cite{shen2012}, which is based on the expected quadratic loss function and modifies Mallows's $C{_p}$ statistics (in linear regression). Given the estimates $\widehat{\bfgamma}=(\widehat{\bfbeta}^\prime, \widehat{\bfrho}^\prime)^{\prime}$ and $\widehat{\bfomega}$, MLIC is calculated by
	\begin{align*}
	MLIC=\sum_{i=1}^{n} (\bfY_i-\widehat{\bfmu}_i)^\prime \bfW_i (\bfY_i-\widehat{\bfmu}_i)+2Tr(\bfE_n^{-1} \bfJ_n),
	\end{align*}
	where $\bfE_n=\sum_{i=1}^{n}\bfD_i^\prime \bfV_i^{-1} \bfW_i \bfD_i$ and $\bfJ_n=\sum_{i=1}^{n}(\bfD_i^\prime \bfV_i^{-1} \bfepsilon_i \bfepsilon_i^\prime -\bfG_i \bfepsilon_i^\prime)\bfD_i$ with $\bfepsilon_i=\bfW_i (\bfY_i-\bfmu_i^{0})$ and $\bfG_i=(\sum_{m=1}^{n}\bfQ_m \bfs_m^\prime)(\sum_{m=1}^{n}\bfs_m \bfs_m^\prime)^{-1} \bfs_i$ where $\bfQ_i=\bfD_i^\prime \bfV_i^{-1} \bfW_i(\bfY_i-\widehat{\bfmu}_i)$ and $\bfs_i$ is the score component of the $i^{th}$ individual in the partial likelihood for the dropout model in (\ref{ees}). {Note that $\bfmu_i^{0}$ is estimated} by the largest candidate model based on the collected information, and numerical studies via simulation have shown that the misspecification of this model has mild or negligible influence on the performance of MLIC. In addition, {\cite{shen2012} also provided} MLICC for correlation structure selection by modifying the penalty term. 
	
	Another commonly used criterion for such context is QICW$_r$ \citep{gosho2016}, which is extended from regular QIC by incorporating the inverse probability weight matrix. Given the estimates $\widehat{\bfgamma}=(\widehat{\bfbeta}^\prime, \widehat{\bfrho}^\prime)^{\prime}$ and $\widehat{\bfomega}$, the QICW$_r$ statistic is provided as
	\begin{align*}
	QICW_r=-2 \sum_{i=1}^{n} \sum_{j=1}^{T} Q_w(\widehat{\bfbeta}, \widehat{\bfomega}; \bfY_{i}, \bfX_{i}, \bfH_{i})+2 Tr(\widehat{\bfPhi}_{I}\widehat{\bfV}_w),
	\end{align*}
	where $Q_w(\widehat{\bfbeta}, \widehat{\bfomega}; \bfY_{ij}, \bfX_{i}, \bfH_{i})$ is the weighted log quasi-likelihood function under an independence correlation structure, and $\widehat{\bfPhi}_{I}=-\sum_{i=1}^{n} \sum_{j=1}^{T}(\partial^2 Q_w/\partial \bfbeta \partial \bfbeta ^\prime) \mid_{\bfbeta=\widehat{\bfbeta}}$. 
	
	%{Both QICW$_r$ and MLIC/MLICC have limited applications, which are referred to Introduction for more details. In this paper, we develop innovative joint selection criteria by embedding empirical likelihood approach into WGEE framework to further improve joint selection performance, which will be described in details next. }
	%However, model selection via empirical likelihood is not straightforward as other likelihood-based methods. The full model is needed to validate this approach.  
	
	\subsubsection{Proposed Criteria of JEAIC and JEBIC}\label{section2.3.2}
	To begin with, we first propose the full weighted estimating equation $\bfG_F$ by accommodating a stationary correlation structure for the empirical likelihood, which is given by
	\begin{align}\label{full}
	\bfG_F\Big(\bfX_{Fi},\bfY_i,\widetilde{\bfbeta},\bfrho^c, \bftheta \Big)=&
	\begin{pmatrix} 
	\bfD_i^\prime \bfV_i^{-1}\bfW_i\big\{\bfY_i-\bfmu_i(\widetilde{\bfbeta})\big\}\\  
	\bfU_i(\widetilde{\bfbeta})-\boldsymbol{h}(\bfrho^c)\phi\\
	\bfs_{i}(\bftheta)
	\end{pmatrix},
	\end{align}
	where $\bfs_i(\bftheta)$ is the estimating equation for $\bftheta$ in (\ref{ees}). Notation $\widetilde{\bfbeta}\in \R^L$ in $\bfG_F$ denotes a vector of parameters with the same dimensionality as $\bfbeta_F\in \R^L$ from our proposed full mean structure with $\bfX_{Fi}$ as the covariates for the $i^{th}$ subject. {Without loss of generality}, we can always rearrange the covariate matrix $\bfX_{Fi}$ so that the first $p-$dimensional vector in $\widetilde{\bfbeta}$ {equals the parameter} vector $\bfbeta$ from the candidate model, and the remaining elements in $\widetilde{\bfbeta}$ {equal zeros}, thus $\widetilde{\bfbeta}=(\bfbeta^\prime, {\bf 0}^\prime)^{\prime}$. In addition, a stationary correlation structure is proposed for the full WGEE to estimate correlation coefficients, i.e., $\bfrho_F^{ST}=(\rho_1^{\text{ST}},\ldots,\rho_{T-1}^{ST})^{\prime}$, $\bfU_i(\widetilde{\bfbeta})=\big(U_{i1}(\widetilde{\bfbeta}), U_{i2}(\widetilde{\bfbeta}),\ldots, U_{i(T-1)}(\widetilde{\bfbeta})\big)^\prime$ with
	%\begin{equation*}
	$U_{im}(\widetilde{\bfbeta})=\sum_{j=1}^{T-m}(R_{i,j+m}/\omega_{i,j+m})e_{ij}(\widetilde{\bfbeta})e_{i,j+m}(\widetilde{\bfbeta}).$
	%\end{equation*}
	Also, for any pre-specified correlation structure denoted by the superscript $c$ (nested within a stationary correlation structure), $\boldsymbol{h}(\bfrho^c)=\Big(\rho_1^c\big(T-1-p/n\big),\ldots,\rho_{T-1}^c(1-p/n)\Big)^\prime$ with $\bfrho^c=(\rho^c_1,...,\rho^c_{T-1})^\prime\in \R^{T-1}$. For instance, $\bfrho^{EXC}=(\rho^{\text{EXC}},...,\rho^{\text{EXC}})^{\prime}$ when {an exchangeable (EXC) correlation structure} is fitted. Here, we consider a stationary correlation structure for the proposed full model; however, it {can} be extended to a more general case (i.e., unstructured), {which may substantially increase the number of parameters needing estimation, and thus likely lead to convergence issues particularly for small $n$ and relatively large $T$}. 
	
	Combining all the information above, we thus have the following empirical likelihood ratio, which is the key component to select marginal mean and correlation structures:
	\begin{equation}\label{el}
	R^F(\bfbeta,\bfrho^c,\bftheta)=\sup_{\bfbeta,\bfrho^c,\bftheta}\left\{ \prod_{i=1}^{n} np_i; p_i>0, \sum_{i=1}^{n}p_i=1, \sum_{i=1}^{n}p_i\bfG_F\big(\bfX_{Fi},\bfY_i,\widetilde{\bfbeta},\bfrho^c, \bftheta\big)=0 \right\},
	\end{equation}
	where $p_i=P(\bfY=\bfy_i,\bfX=\bfx_i)$. Here, we assume that only the distributions with an atom of probability on each $\bfy_i$ and $\bfx_i$ have nonzero likelihood. {Therefore, $\{p_i\}$'s will follow the rule of traditional probability with the sum equal to one.} {Without imposing constraints defined by the estimating equations}, $\prod_{i=1}^{n} p_i$ is maximized as $\prod_{i=1}^{n} (1/n)$. Thus, the empirical likelihood ratio is defined as $\prod_{i=1}^{n} np_i$. {More basic properties about empirical likelihood can be found in \cite{owen2001}}. An intuitive rationale of model selection based on proposed empirical likelihood ratio is as follows: when the estimators $\widehat{\bfbeta}_F$, $\widehat{\bfrho}^{ST}_F=(\widehat{\rho}_1^{\text{ST}},...,\widehat{\rho}_{T-1}^{ST})^{\prime}$ are obtained from {the WGEE method} with $\bfX_{Fi}$ and a stationary correlation structure from (\ref{wgee}), and $\widehat{\bftheta}$ is calculated from (\ref{ees}), we will have $R^F(\widehat{\bfbeta}_F, \widehat{\bfrho}^{ST}_F, \widehat{\bftheta})=1$, which achieves the upper limit of the empirical likelihood ratio. However, the estimators $\widehat{\bfbeta}$ and $\widehat{\bfrho}^c$ other than $\widehat{\bfbeta}_F$ and $\widehat{\bfrho}^{ST}_F$ {will lead to} $R^F(\widehat{\bfbeta}, \widehat{\bfrho}^c, \widehat{\bftheta})<1$. The departure from 1 indicates the misspecification of the model to the degree reflected by the magnitude of the deviation. In other words, {the closer the mean and correlation structures approach the underlying true values, the closer $R^F$ will approach $1$}, which ensures the potential for joint selection of marginal mean and correlation structures. 
	
	Thereafter, by plugging the parameter estimates $(\widehat{\bfbeta}^\prime,\widehat{\bfrho}^{c\prime})^\prime$ from a candidate model in WGEE (\ref{wgee}) and $\widehat{\bftheta}_{ML}$ obtained based on the estimating equation (\ref{ees}) into $R^F(\widehat{\bfbeta},\widehat{\bfrho}^c, \widehat{\bftheta}_{ML})$, the empirical likelihood ratio is {the solution of the following equation by utilizing the Lagrange multiplier method \citep{owen2001},} 
	\begin{equation}\label{elr}
	-2\log R^F(\widehat{\bfbeta},\widehat{\bfrho}^c,\widehat{\bftheta}_{ML})=2\sum_{i=1}^{n}\log\big\{1+\boldsymbol{\lambda}^\prime\bfG_F(\bfX_i,\bfY_i,\widehat{\widetilde{\bfbeta}},\widehat{\bfrho}^c, \widehat{\bftheta}_{ML})\big\},
	\end{equation}
	where the parameter $\boldsymbol{\lambda}$ can be solved by applying the Newton-Raphson method based on
	\begin{equation}\label{lambda}
	\sum_{i=1}^{n}\frac{\bfG_F(\bfX_i,\bfY_i,\widehat{\widetilde{\bfbeta}},\widehat{\bfrho}^c, \widehat{\bftheta}_{ML})}{1+\boldsymbol{\lambda}^\prime\bfG_F(\bfX_i,\bfY_i,\widehat{\widetilde{\bfbeta}},\widehat{\bfrho}^c, \widehat{\bftheta}_{ML})}=0.
	\end{equation}
	
	Thus, for longitudinal data with dropout missingness under MAR, our proposed information criteria are defined by
	\begin{equation*}
	\begin{split}
	\text{JEAIC}&=-2\log R^F(\widehat{\bfbeta},\widehat{\bfrho}^c, \widehat{\bftheta}_{ML})+2\widetilde{p},\\
	\text{JEBIC}&=-2\log R^F(\widehat{\bfbeta},\widehat{\bfrho}^c, \widehat{\bftheta}_{ML})+\widetilde{p}\log n,  
	\end{split}
	\end{equation*}
	where $\widetilde{p}$ denotes the total number of parameters. The asymptotic property of our proposed information criteria can be evaluated based on the existing work. In particular, in the work by \cite{eic1995}, EIC has been proved to be an asymptotically unbiased estimate that is proportional to the expected Kullback-Leibler distance {between two discrete empirical distributions}. Also, \cite{variyath2010} evaluated the consistency of EBIC. {In both of their works, general estimating equations are considered, but it is straightforward to embed our proposed full estimating equations (\ref{full}) into their theoretical framework when the empirical likelihood estimators are utilized.} {However, our proposed approach is built upon the plug-in estimators, thus, it is important to assess the asymptotic proprieties of these plug-in estimators and their relationship with the empirical likelihood estimators. } 
	%before achieving this, the challenge is to explore the asymptotic association of these plug-in estimators and empirical-likelihood-based estimators, which will be provided next with detailed proofs. %In addition, the pseudo algorithm procedures for joint selection of marginal mean and correlation structures are summarized here: {Step I: calculate $\widehat{\bftheta}_{ML}$ based on (\ref{ees}); Step II: obtain $\widehat{\bfbeta}$ and $\widehat{\bfrho}$ from a pre-specified WGEE candidate model given $\widehat{\bftheta}_{ML}$; Step III: calculate $R^F(\widehat{\bfbeta},\widehat{\bfrho}^c; \widehat{\bftheta}_{ML})$ to get JEAIC (JEBIC); Step IV, Repeat Step I and Step II to get JEAIC (JEBIC) for all candidate models, and select the mean and correlation structures which minimize JEAIC (JEBIC).} 
	
	%in another word, EAIC might select false positive predictors with probability asymptotically larger than zero, while EBIC will falsely select significant predictors with probability tending to zero. 
	\subsubsection{Asymptotic Properties of Plug-in Estimators}
	In this section, we will investigate the asymptotic properties of our plug-in estimators under MAR, and explain why we advocate {such an alternative}. First, we investigate the asymptotic behavior of estimators $\widehat{\bfbeta}_{EL}$, $\widehat{\bfrho}^c_{EL}$, and $\widehat{\bftheta}_{EL}$ from maximizing the profile empirical likelihood ratio. Inspired by \cite{qin1994} and \cite{qin2009}, we derive the asymptotic properties of the estimator shown in Theorem \ref{theorem 1} with the proof sketched in {the Supporting Information}.
	\begin{theorem}\label{theorem 1}
		Let us denote 
		\begin{equation*}
		\begin{split}
		&\bfg_F(\bfX_i, \bfY_i, \widetilde{\bfbeta},\bfrho^c, \bftheta)=
		\begin{pmatrix} 
		\bfD_i^\prime \bfV_i^{-1}\bfW_i\Big\{\bfY_i-\bfmu_i(\widetilde{\bfbeta})\Big\}\\  
		\bfU_i(\bfbeta)-\boldsymbol{h}(\bfrho^c)\phi
		\end{pmatrix}, \\
		&\big(\widehat{\bfbeta}^\prime_{EL}, \widehat{\bfrho}^{c\prime}_{EL}, \widehat{\bftheta}^\prime_{EL}\big)^\prime    =\argmax_{\bfbeta,\bfrho^c,\bftheta} R^F(\bfbeta,\bfrho^c,\bftheta), ~\text{and}~ \bfs=\bfs_i(\bftheta). 
		\end{split}
		\end{equation*}
		Under the conditions specified in {the Supporting Information} and given $\bfgamma=(\bfbeta^\prime,\bfrho^{c\prime})^\prime$ and $\bftheta$ with corresponding true values $\bfgamma_0$ and $\bftheta_0$, we have
		\begin{enumerate}[(1)]
			\item
			\begin{equation}\label{taylor}
			\left(\begin{array}{c}
			\widehat{\bfgamma}_{EL}-\bfgamma_0\\ \widehat{\bftheta}_{EL}-\bftheta_0    \end{array}\right)=
			\left(\begin{array}{c}
			-\bfV_{\ast}\bfA_{\ast}\bfQ_{n}^\ast\\ \bfOmega\bfS_{n\bftheta}    \end{array}\right)
			+o_p(\textbf{n}^{-\frac{1}{2}}),
			\end{equation}
			where $\bfS_{n\bftheta}$ is defined in (\ref{ees}), and 
			\begin{equation*}
			\begin{split}
			\bfV_\ast=&\Bigg[E\big(\frac{\partial \bfg_{F} }{\partial\bfgamma^\prime} \big)^\prime \Big\{E\bfg_{F}\bfg_{F}^\prime-E\big(\frac{\partial \bfg_{F} }{\partial\bftheta^\prime} \big) \big(E\bfs\bfs^\prime\big)^{-1}E\big(\frac{\partial \bfg_F  }{\partial\bftheta^\prime}\big)^\prime\Big\}^{-1}E\big(\frac{\partial \bfg_F  }{\partial\bfgamma^\prime} \big)\Bigg]^{-1},\\
			\bfA_\ast=&E\big(\frac{\partial \bfg_{F} }{\partial\bfgamma^\prime} \big)^\prime \Big\{E\bfg_{F}\bfg_{F}^\prime-E\big(\frac{\partial \bfg_{F} }{\partial\bftheta^\prime} \big) \big(E\bfs\bfs^\prime\big)^{-1}E\big(\frac{\partial \bfg_F  }{\partial\bftheta^\prime}\big)^\prime\Big\}^{-1}, \\
			\bfQ^\ast_{n}=&\frac{1}{n} \sum_{i=1}^{n} \bfg_F(\bfX_i, \bfY_i, \widetilde{\bfbeta},\bfrho^c, \bftheta)+E\big(\frac{\partial \bfg_F }{\partial\bftheta^\prime} \big)^\prime E(\bfs\bfs^\prime)^{-1}\bfS_{n\bftheta}, ~ \bfOmega=\big(E\bfs\bfs^\prime\big)^{-1}.
			\end{split}
			\end{equation*}
			
			\item Furthermore, the asymptotic normality can be derived from (\ref{taylor})
			\begin{equation*}
			\sqrt{n}\left(\begin{array}{c}
			\widehat{\bfgamma}_{EL}-\bfgamma_0\\ \widehat{\bftheta}_{EL}-\bftheta_0 \end{array}\right) \overset{d}{\to}
			\bfN\left(\left(\begin{array}{c}
			\bf0\\ \bf0    \end{array}\right), \left(\begin{array}{cc}
			\bfSigma_{11} & \bf0\\ \bf0 & \bfSigma_{22}    \end{array}\right)\right),
			\end{equation*}
			with $\bfSigma_{11}=\bfV_{\ast}\bfA_{\ast}\textbf{Cov}(\bfQ_{n}^\ast)\bfA_{\ast}^\prime\bfV_{\ast}^\prime$, $\bfSigma_{22}=\bfOmega\textbf{Cov}(\bfS_{n\bftheta})\bfOmega^\prime$.
			
			\item {$-2\log R^F(\widehat{\bfbeta}_{EL}, \widehat{\bfrho}^c_{EL}, \widehat{\bftheta}_{EL})$ follows a $\chi^2$ distribution with $\widetilde{L}-\widetilde{p}$ degrees of freedom where $\widetilde{L}$ is the number of estimating equations in (\ref{full})} and $\widetilde{p}$ as the total number of parameters.    
		\end{enumerate}
	\end{theorem}
	
	An interesting finding from Theorem \ref{theorem 1} is that the empirical-likelihood-based estimator $\widehat{\bftheta}_{EL}$ is asymptotically equivalent to the estimator $\widehat{\bftheta}_{ML}$ from partial likelihood in (\ref{ees}) since they have the same influence function. Also, the estimator $\widehat{\bftheta}_{EL}$ {is asymptotically independent of} the estimator $\widehat{\bfgamma}_{EL}$ by Theorem \ref{theorem 1} (II). Thus, we can substitute $\widehat{\bftheta}_{ML}$ {in} $R^F(\bfbeta,\bfrho^c,\bftheta)$ first and then estimate $\bfgamma$ {by maximizing} $R^F(\bfgamma;\widehat{\bftheta}_{ML})$, {by which means, the estimator is asymptotically equivalent to the estimator $\widehat{\bfgamma}_{EL}$, thus we keep this notation for this context.} Such plug-in method can definitely decrease the dimensionality of parameters for estimation by only focusing on $\bfgamma$, and thus {reducing the computational} burden in particular when the dimension of $\bftheta$ is relatively large.
	
	However, maximizing $R^F(\bfgamma;\widehat{\bftheta}_{ML})$ to estimate $\bfgamma$ still raises {computational issues} since the number of the estimating equations may exceed the number of parameters, which requires 0 to be inside the convex hull of data to guarantee the existence of solution \citep{chen2012, variyath2010}. Furthermore, the bounded support of correlation coefficients also increases the difficulty among the existing algorithms. Instead, we advocate to substitute the empirical likelihood estimators $\widehat{\bfgamma}_{EL}$ in $R^F(\widehat{\bfgamma}_{EL}; \widehat{\bftheta}_{ML})$ with the estimators {from a candidate model fitting in WGEE} (\ref{wgee}), which can {avoid computational issues} and ensure convenient application. Here, we investigate the asymptotic relationship between the WGEE and empirical-likelihood-based estimators, which is summarized in the following theorem: 
	\begin{theorem}\label{theorem 2}
		Under Theorem \ref{theorem 1} and the conditions provided in {the Supporting Information}, the estimates $\widehat{\bfgamma}_{EL}=(\widehat{\bfbeta}^\prime_{EL},\widehat{\bfrho}^{c\prime}_{EL})^\prime$ from empirical likelihood based on (\ref{el}) and $\widehat{\bfgamma}=(\widehat{\bfbeta}^\prime,\widehat{\bfrho}^{c\prime})^{\prime}$ based on WGEE (\ref{wgee}) are asymptotically equivalent.
	\end{theorem}
	
	The proofs for exchangeable and AR1 scenarios are provided in {the Supporting Information}. Theorem \ref{theorem 2} implies that the WGEE estimator is a reasonable approximation of the empirical likelihood estimator under certain conditions, indicating that any asymptotic properties induced by the empirical likelihood estimator would be reasonably invoked by the WGEE estimator. More discussion on conditions is referred to {the Supporting Information}.
	
	\section{Simulation studies}
	\label{s:simulation}
	
	In this section, we investigate the numerical performance of our proposed criteria under various settings,
	%We further release our assumption to the case where overdispersion is unknown, which can be estimated in WGEE procedure:
	%\begin{equation*}
	%    \widehat{\phi}(\widehat{\bfbeta})=\frac{1}{nT-p}\sum_{i=1}^{n}\sum_{j=1}^{T}\Big(\frac{R_{ij}}{\widehat{\omega}_{ij}}\Big)e^2_{ij}(\widehat{\bfbeta}).
	%\end{equation*}
	and compare with several existing criteria such as MLIC and QICW$_r$ as well as the most recent work of JLIC. We expect better performance of the two proposed criteria compared to the existing alternatives. {In addition}, JEBIC might have better control of false positive rates than JEAIC under relatively large sample sizes \citep{variyath2010}.
	
	Our first scenario considers binary outcomes, and the true marginal mean structure is
	\begin{equation}
	\log\big(\frac{\mu_{ij}}{1-\mu_{ij}}\big)=\beta_0+x_{i1}\beta_1+x_{ij2}\beta_2, ~\text{for } i=1,...,n, j=1,...,T,
	\end{equation}
	where $x_{i1}$ is the subject (cluster) level covariate generated from {the uniform distribution over $[0,1]$} and $x_{ij2}=j-1$ is a time-dependent covariate. The number of observations (i.e., cluster size) is $T=3$. The true parameter vector $\bfbeta=(\beta_0, \beta_1, \beta_2)^{\prime}$ in the marginal mean is $(-1, 1, 0.4)^{\prime}$. The true correlation structure is exchangeable with a correlation coefficient $\rho_0=0.5$. The dropout model is 
	\begin{equation}
	\log\big(\frac{\lambda_{ij}}{1-\lambda_{ij}}\big)=\theta_0+y_{i(j-1)}\theta_1+h_{ij}\theta_2, ~\text{for } i=1,...,n, j=2,...,T,
	\end{equation}
	{where the covariate $h_{ij}$ is uniformly distributed over $[-0.5,0.5]$}. Different choices for the parameters $\bftheta=(\theta_0, \theta_1, \theta_2)^{\prime}$ can ensure the missing probability (denoted by $m$) around $0.2$ and $0.3$, i.e., $\bftheta=(1.74, 0.5, -0.8)^{\prime}$ is for $m=0.2$ and $\bftheta=(1.05, 0.5, -0.8)^{\prime}$ is for $m=0.3$. 
	
	{In the first scenario, we consider a correctly specified dropout model.} Then, we also evaluate the robustness of our proposal when {the dropout model is misspecified because of the left out variable $h_{ij}$ in the regression \citep{shen2017}.}
	%\begin{equation*}
	%\log\big(\frac{\lambda_{ij}}{1-\lambda_{ij}}\big)=\theta_0+y_{i(j-1)}\theta_1, ~\text{for } i=1,...,n, j=2,...,T.
	%\end{equation*}
	
	In addition, we generate one redundant variable {$x_{ij3} \sim N(0,1)$}. The full model considered for our proposed criteria as well as MLIC/MLICC includes three variables, $x_{i1}$, $x_{ij2}$ and $x_{ij3}$. Six potential marginal mean structures are considered with three types of ``working" correlation structures (i.e., exchangeable (EXC), AR1 and Independence (IND)) for model fitting. To {summarize the simulation results}, 500 Monte Carlo data sets with sample size $n=100, 200$ are generated for each scenario, and the selection rate for each combination of marginal mean and correlation structures is reported. Moreover, we also consider the scenarios with Gaussian outcomes, the ones {where} the assumption of MAR is violated, and also the ones with redundant variables. Due to limited space, we cannot show all these results here, but provide them in {the Supporting Information}.
	
	{On the other hand}, to compare our proposal with JLIC, we consider the same set-ups (with binary and Gaussian outcomes) in \cite{shen2017} by utilizing their supporting program functions for simulations. The detailed information on parameter set-ups is not provided here but can be referred to \cite{shen2017}. All the simulations are conducted in R and MATLAB software.
	\begin{table}[ht]
		\centering
		\footnotesize
		\begin{minipage}{160mm}
			\caption{Performance of JEAIC and JEBIC compared with MLIC and QICW$_r$: Percentage of selecting six candidate logistic models across 500 Monte Carlo datasets; $T=3$, $\rho=0.5$. The model with $\{x_1,x_2\}$ and {an EXC} correlation structure is the true model. Notation $n$ and $m$ {denote the sample size and the missing probability}, respectively.}
			\label{table1}
			\rowcolors{2}{gray!25}{white}
			\begin{tabular}{p{1cm} p{1.2cm} p{1.2cm} p{1.2cm} p{1.2cm} p{1.2cm} p{1.2cm} p{1.2cm} p{1.2cm} p{1.2cm}}
				\rowcolor{gray!50}
				\toprule
				Setups &Method & $\bfC(\bfrho)$ & $x_1$ & $x_3$ & $\bfx_1,\bfx_2$ & $x_1,x_3$ & $x_2,x_3$ & $x_1,x_2,x_3$ & Total \\  \midrule
				n=100  & JEAIC & AR1 & 0.004 & 0 & 0.082 & 0 & 0.016 & 0.006 & 0.108 \\
				m=0.2 & & \textbf{EXC} & 0.026 & 0.008 & \textbf{0.578} & 0.002 & 0.186 & 0.092 & \textbf{0.892} \\
				& & IND & 0 & 0 & 0 & 0 & 0 & 0 & 0 \\
				& & Total & 0.03 & 0.008 & \textbf{0.66} & 0.002 & 0.202 & 0.098 & 1 \\
				& JEBIC & AR1 & 0.02 & 0.004 & 0.072 & 0 & 0.014 & 0 & 0.11 \\
				& & \textbf{EXC} & 0.09 & 0.028 & \textbf{0.566} & 0.002 & 0.2 & 0.004 & \textbf{0.89} \\
				& & IND & 0 & 0 & 0 & 0 & 0 & 0 & 0 \\
				& & Total & 0.11 & 0.032 & \textbf{0.638} & 0.002 & 0.214 & 0.004 & 1 \\
				& MLIC & AR1 & 0.008 & 0.008 & 0.2 & 0.002 & 0.14 & 0.06 & 0.418 \\
				& & \textbf{EXC} & 0.008 & 0.008 & \textbf{0.28} & 0.004 & 0.168 & 0.068 & \textbf{0.536} \\
				& & IND & 0.004 & 0 & 0.018 & 0 & 0.016 & 0.008 & 0.046 \\
				& & Total & 0.02 & 0.016 & \textbf{0.498} & 0.006 & 0.324 & 0.136 & 1 \\
				& QICW$_r$ & AR1 & 0 & 0 & 0.062 & 0 & 0.038 & 0.04 & 0.14 \\
				& & \textbf{EXC} & 0.006 & 0.004 & \textbf{0.436} & 0.002 & 0.236 & 0.112 & \textbf{0.796} \\
				& & IND & 0 & 0 & 0.03 & 0 & 0.02 & 0.014 & 0.064 \\
				& & Total & 0.006 & 0.004 & \textbf{0.528} & 0.002 & 0.294 & 0.166 & 1 \\ \midrule
				n=100  & JEAIC & AR1 & 0.01 & 0.002 & 0.102 & 0 & 0.03 & 0.016 & 0.16 \\
				m=0.3 & & \textbf{EXC }& 0.042 & 0.026 & \textbf{0.472} & 0.004 & 0.198 & 0.098 & \textbf{0.84} \\
				& & IND & 0 & 0 & 0 & 0 & 0 & 0 & 0 \\
				& & Total & 0.052 & 0.028 & \textbf{0.574} & 0.004 & 0.228 & 0.114 & 1 \\
				& JEBIC & AR1 & 0.038 & 0.014 & 0.082 & 0 & 0.028 & 0.002 & 0.164 \\
				& & \textbf{EXC }& 0.126 & 0.066 & \textbf{0.44} & 0.002 & 0.188 & 0.014 & \textbf{0.836} \\
				& & IND & 0 & 0 & 0 & 0 & 0 & 0 & 0 \\
				& & Total & 0.164 & 0.08 & \textbf{0.522} & 0.002 & 0.216 & 0.016 & 1 \\
				& MLIC & AR1 & 0.01 & 0.01 & 0.164 & 0 & 0.106 & 0.064 & 0.354 \\
				& & \textbf{EXC} & 0.036 & 0.026 & \textbf{0.29} & 0.002 & 0.174 & 0.06 & \textbf{0.588} \\
				& & IND & 0.002 & 0.004 & 0.028 & 0.002 & 0.014 & 0.008 & 0.058 \\
				& & Total & 0.048 & 0.04 & \textbf{0.482} & 0.004 & 0.294 & 0.132 & 1 \\
				& QICW$_r$ & AR1 & 0.002 & 0.002 & 0.05 & 0.002 & 0.028 & 0.03 & 0.114 \\
				& & \textbf{EXC} & 0.008 & 0.006 & \textbf{0.452} & 0.002 & 0.232 & 0.136 & \textbf{0.836} \\
				& & IND & 0 & 0 & 0.026 & 0 & 0.012 & 0.012 & 0.05 \\
				& & Total & 0.01 & 0.008 & \textbf{0.528} & 0.004 & 0.272 & 0.178 & 1 \\ \midrule
				n=200 & JEAIC & AR1 & 0 & 0 & 0.034 & 0 & 0.008 & 0.008 & 0.05 \\
				m=0.2 & & \textbf{EXC} & 0 & 0 & \textbf{0.73} & 0 & 0.096 & 0.124 & \textbf{0.95} \\
				& & IND & 0 & 0 & 0 & 0 & 0 & 0 & 0 \\
				& & Total & 0 & 0 & \textbf{0.764} & 0 & 0.104 & 0.132 & 1 \\
				& JEBIC & AR1 & 0 & 0 & 0.042 & 0 & 0.012 & 0 & 0.054 \\
				& & \textbf{EXC }& 0.01 & 0 & \textbf{0.806} & 0 & 0.114 & 0.016 & \textbf{0.946} \\
				& & IND & 0 & 0 & 0 & 0 & 0 & 0 & 0 \\
				& & Total & 0.01 & 0 & \textbf{0.848} & 0 & 0.126 & 0.016 & 1 \\
				& MLIC & AR1 & 0 & 0 & 0.23 & 0 & 0.064 & 0.068 & 0.362 \\
				& & \textbf{EXC} & 0.002 & 0 & \textbf{0.392} & 0 & 0.082 & 0.098 & \textbf{0.574} \\
				& & IND & 0.002 & 0 & 0.036 & 0 & 0.006 & 0.02 & 0.064 \\
				& & Total & 0.004 & 0 & \textbf{0.658} & 0 & 0.152 & 0.186 & 1 \\
				& QICW$_r$ & AR1 & 0 & 0 & 0.056 & 0 & 0.012 & 0.02 & 0.088 \\
				& & \textbf{EXC} & 0 & 0 & \textbf{0.56} & 0 & 0.114 & 0.168 & \textbf{0.842} \\
				& & IND & 0 & 0 & 0.04 & 0 & 0.002 & 0.028 & 0.07 \\
				& & Total & 0 & 0 & \textbf{0.656} & 0 & 0.128 & 0.216 & 1 \\ \midrule
				n=200 & JEAIC & AR1 & 0 & 0 & 0.066 & 0 & 0.014 & 0.008 & 0.088 \\
				m=0.3 & & \textbf{EXC} & 0.006 & 0 & \textbf{0.646} & 0 & 0.132 & 0.128 & \textbf{0.912} \\
				& & IND & 0 & 0 & 0 & 0 & 0 & 0 & 0 \\
				& & Total & 0.006 & 0 & \textbf{0.712} & 0 & 0.146 & 0.136 & 1 \\
				& JEBIC & AR1 & 0.002 & 0 & 0.074 & 0 & 0.014 & 0 & 0.09 \\
				& & \textbf{EXC} & 0.038 & 0.004 & \textbf{0.704} & 0.002 & 0.152 & 0.01 & \textbf{0.91} \\
				& & IND & 0 & 0 & 0 & 0 & 0 & 0 & 0 \\
				& & Total & 0.04 & 0.004 & \textbf{0.778} & 0.002 & 0.166 & 0.01 & 1 \\
				& MLIC & AR1 & 0.002 & 0 & 0.214 & 0.002 & 0.056 & 0.056 & 0.33 \\
				& & \textbf{EXC} & 0.002 & 0.002 & \textbf{0.386} & 0 & 0.124 & 0.098 & \textbf{0.612} \\
				& & IND & 0 & 0 & 0.03 & 0 & 0.01 & 0.018 & 0.058 \\
				& & Total & 0.004 & 0.002 & \textbf{0.63} & 0.002 & 0.19 & 0.172 & 1 \\
				& QICW$_r$ & AR1 & 0 & 0 & 0.066 & 0 & 0.006 & 0.03 & 0.102 \\
				& & \textbf{EXC} & 0 & 0 &\textbf{0.554} & 0 & 0.118 & 0.18 & \textbf{0.852} \\
				& & IND & 0 & 0 & 0.018 & 0 & 0.004 & 0.024 & 0.046 \\
				& & Total & 0 & 0 & \textbf{0.638} & 0 & 0.128 & 0.234 & 1 \\
				\bottomrule
			\end{tabular}
		\end{minipage}
		\vspace*{-6pt}
	\end{table}
	
	\begin{table}[ht]
		\centering
		\footnotesize
		\begin{minipage}{160mm}
			\caption{Performance of JEAIC and JEBIC compared with MLIC and QICW$_r$ when the dropout model is misspecified: Percentage of selecting six candidate logistic models across 500 Monte Carlo datasets; $T=3$, $\rho=0.3$. The model with $\{x_1,x_2\}$ and {an EXC} correlation structure is the true model. Notation $n$ and $m$ denote {denote the sample size and the missing probability}, respectively.}
			\label{mis_dropout}
			\label{table2}
			\rowcolors{2}{gray!25}{white}
			\begin{tabular}{p{1cm} p{1.2cm} p{1.2cm} p{1.2cm} p{1.2cm} p{1.2cm} p{1.2cm} p{1.2cm} p{1.2cm} p{1.2cm}}
				\rowcolor{gray!50}
				\toprule
				Setups &Method & $\bfC(\bfrho)$ & $x_1$ & $x_3$ & $\bfx_1,\bfx_2$ & $x_1,x_3$ & $x_2,x_3$ & $x_1,x_2,x_3$ & Total \\  \midrule
				n=100& JEAIC & AR1          & 0.002 & 0.002 & 0.092          & 0     & 0.012 & 0.008 & 0.116          \\
				m=0.2  &       & \textbf{EXC} & 0.024 & 0.01  & \textbf{0.566} & 0.002 & 0.191 & 0.09  & \textbf{0.884} \\
				&       & IND          & 0     & 0     & 0              & 0     & 0     & 0     & 0              \\
				&       & Total        & 0.026 & 0.012 & \textbf{0.659} & 0.002 & 0.203 & 0.098 & 1              \\
				& JEBIC & AR1          & 0.022 & 0.004 & 0.084          & 0     & 0.012 & 0     & 0.122          \\
				&       & \textbf{EXC} & 0.096 & 0.022 & \textbf{0.557} & 0.002 & 0.195 & 0.006 & \textbf{0.878} \\
				&       & IND          & 0     & 0     & 0              & 0     & 0     & 0     & 0              \\
				&       & Total        & 0.118 & 0.026 & \textbf{0.641} & 0.002 & 0.207 & 0.006 & 1              \\
				& MLIC  & AR1          & 0.006 & 0.012 & 0.212          & 0     & 0.126 & 0.054 & 0.41           \\
				&       & \textbf{EXC} & 0.012 & 0.006 & \textbf{0.258} & 0.006 & 0.184 & 0.074 & \textbf{0.54}  \\
				&       & IND          & 0     & 0     & 0.028          & 0     & 0.016 & 0.006 & 0.05           \\
				&       & Total        & 0.018 & 0.018 & \textbf{0.498} & 0.006 & 0.326 & 0.134 & 1              \\
				& QICW$_r$  & AR1          & 0     & 0     & 0.056          & 0     & 0.042 & 0.04  & 0.138          \\
				&       & \textbf{EXC} & 0.008 & 0.004 & \textbf{0.436} & 0.002 & 0.232 & 0.114 & \textbf{0.796} \\
				&       & IND          & 0     & 0     & 0.034          & 0     & 0.02  & 0.012 & 0.066          \\
				&       & Total        & 0.008 & 0.004 & \textbf{0.526} & 0.002 & 0.294 & 0.166 & 1              \\\midrule
				n=100 & JEAIC & AR1          & 0.01  & 0.002 & 0.094          & 0.002 & 0.032 & 0.02  & 0.16           \\
				m=0.3 &       & \textbf{EXC} & 0.046 & 0.026 & \textbf{0.484} & 0.004 & 0.194 & 0.086 & \textbf{0.84}  \\
				&       & IND          & 0     & 0     & 0              & 0     & 0     & 0     & 0              \\
				&       & Total        & 0.056 & 0.028 & \textbf{0.578} & 0.006 & 0.226 & 0.106 & 1              \\
				& JEBIC & AR1          & 0.042 & 0.012 & 0.078          & 0     & 0.028 & 0.002 & 0.162          \\
				&       & \textbf{EXC} & 0.142 & 0.066 & \textbf{0.436} & 0     & 0.184 & 0.01  & \textbf{0.838} \\
				&       & IND          & 0     & 0     & 0              & 0     & 0     & 0     & 0              \\
				&       & Total        & 0.184 & 0.078 & \textbf{0.514} & 0     & 0.212 & 0.012 & 1              \\
				& MLIC  & AR1          & 0.01  & 0.008 & 0.154          & 0.002 & 0.098 & 0.046 & 0.318          \\
				&       & \textbf{EXC} & 0.038 & 0.032 & \textbf{0.296} & 0.002 & 0.184 & 0.066 & \textbf{0.618} \\
				&       & IND          & 0.002 & 0.004 & 0.028          & 0     & 0.016 & 0.014 & 0.064          \\
				&       & Total        & 0.05  & 0.044 & \textbf{0.478} & 0.004 & 0.298 & 0.126 & 1              \\
				& QICW$_r$  & AR1          & 0.002 & 0     & 0.048          & 0     & 0.03  & 0.034 & 0.114          \\
				&       & \textbf{EXC} & 0.008 & 0.004 & \textbf{0.456} & 0.004 & 0.232 & 0.132 & \textbf{0.836} \\
				&       & IND          & 0     & 0     & 0.022          & 0     & 0.012 & 0.016 & 0.05           \\
				&       & Total        & 0.01  & 0.004 & \textbf{0.526} & 0.004 & 0.274 & 0.182 & 1          \\ \midrule
				n=200 & JEAIC & AR1          & 0     & 0     & 0.036          & 0     & 0.008 & 0.008 & 0.052          \\
				m=0.2 	&       & \textbf{EXC} & 0     & 0     & \textbf{0.726} & 0     & 0.092 & 0.13  & \textbf{0.948} \\
				&       & IND          & 0     & 0     & 0              & 0     & 0     & 0     & 0              \\
				&       & Total        & 0     & 0     & \textbf{0.762} & 0     & 0.1   & 0.138 & 1              \\
				& JEBIC & AR1          & 0     & 0     & 0.042          & 0     & 0.012 & 0.002 & 0.056          \\
				&       & \textbf{EXC} & 0.01  & 0     & \textbf{0.806} & 0     & 0.11  & 0.018 & \textbf{0.944} \\
				&       & IND          & 0     & 0     & 0              & 0     & 0     & 0     & 0              \\
				&       & Total        & 0.01  & 0     & \textbf{0.848} & 0     & 0.122 & 0.02  & 1              \\
				& MLIC  & AR1          & 0     & 0     & 0.216          & 0     & 0.05  & 0.072 & 0.338          \\
				&       & \textbf{EXC} & 0     & 0     & \textbf{0.404} & 0     & 0.098 & 0.098 & \textbf{0.6}   \\
				&       & IND          & 0.002 & 0     & 0.032          & 0     & 0.006 & 0.022 & 0.062          \\
				&       & Total        & 0.002 & 0     & \textbf{0.652} & 0     & 0.154 & 0.192 & 1              \\
				& QICW$_r$ & AR1          & 0     & 0     & 0.056          & 0     & 0.01  & 0.028 & 0.094          \\
				&       & \textbf{EXC} & 0     & 0     & \textbf{0.558} & 0     & 0.108 & 0.168 & \textbf{0.834} \\
				&       & IND          & 0     & 0     & 0.04           & 0     & 0.002 & 0.03  & 0.072          \\
				&       & Total        & 0     & 0     & \textbf{0.654} & 0     & 0.12  & 0.226 & 1              \\ \midrule
				n=200 & JEAIC & AR1          & 0     & 0     & 0.068          & 0     & 0.012 & 0.008 & 0.088          \\
				m=0.3 	&       & \textbf{EXC} & 0.008 & 0     & \textbf{0.662} & 0     & 0.132 & 0.11  & \textbf{0.912} \\
				&       & IND          & 0     & 0     & 0              & 0     & 0     & 0     & 0              \\
				&       & Total        & 0.008 & 0     & 0.73           & 0     & 0.144 & 0.118 & 1              \\
				& JEBIC & AR1          & 0.004 & 0.002 & 0.078          & 0     & 0.012 & 0     & 0.096          \\
				&       & \textbf{EXC} & 0.04  & 0     & \textbf{0.698} & 0.002 & 0.154 & 0.01  & \textbf{0.904} \\
				&       & IND          & 0     & 0     & 0              & 0     & 0     & 0     & 0              \\
				&       & Total        & 0.044 & 0.002 & \textbf{0.776} & 0.002 & 0.166 & 0.01  & 1              \\
				& MLIC  & AR1          & 0.002 & 0     & 0.206          & 0.002 & 0.054 & 0.072 & 0.336          \\
				&       & \textbf{EXC} & 0.004 & 0.002 & \textbf{0.374} & 0     & 0.13  & 0.094 & \textbf{0.604} \\
				&       & IND          & 0     & 0     & 0.032          & 0     & 0.012 & 0.016 & 0.06           \\
				&       & Total        & 0.006 & 0.002 & \textbf{0.612} & 0.002 & 0.196 & 0.182 & 1              \\
				& QICW$_r$  & AR1          & 0     & 0     & 0.064          & 0     & 0.004 & 0.028 & 0.096          \\
				&       & \textbf{EXC} & 0     & 0     & \textbf{0.542} & 0     & 0.124 & 0.192 & \textbf{0.858} \\
				&       & IND          & 0     & 0     & 0.018          & 0     & 0.004 & 0.024 & 0.046          \\
				&       & Total        & 0     & 0     & \textbf{0.624} & 0     & 0.132 & 0.244 & 1             \\
				\bottomrule
			\end{tabular}
		\end{minipage}
		\vspace*{-6pt}
	\end{table}
	
	In Table \ref{table1}, We find out that both JEAIC and JEBIC outperform two-stage MLIC/MLICC and QICW$_r$ across different settings. In general, all methods exhibit better selection behaviors if sample size increases or missing probability decreases, but the superiority of our proposal becomes more apparent compared to the other alternatives regarding higher improvement in selection rates. Under relatively small sample size, JEAIC and JEBIC behave similarly on joint model selection, while JEBIC seems more promising under relatively large sample size by imposing more penalty on both parameter number and sample size, which agrees with our expectation \citep{variyath2010}. On the other hand, the performances of MLIC/MLICC and QICW$_r$ are not satisfactory and consistently stable across different setups despite having slightly better performance as the sample size increases. Similar patterns and selection rates can be found in Table \ref{table2} {, which indicates} that misspecification of {the dropout model} does not have much influence on the performance of our proposed criteria when the MAR assumption still holds. 
	
	Moreover, using the same set-ups in the first scenario, we conduct further investigation by only considering marginal mean selection given a pre-specified correlation structure according to the editor's suggestion. The results, in {the Supporting Information}, imply that the misspecified correlation structure would worsen the selection performance. More interestingly, in Table \ref{table1}, the marginal selection rates, for mean structures (column total) regardless of the correlation structure selection, is comparable or even slightly higher than the Oracle one under which the true correlation structure is specified and fixed for the marginal mean selection. These findings provide further evidence of our joint selection's advantages; thus, even though the marginal mean structure is the sole interest, the implementation of the joint selection would promise a satisfactory selection rate. Also, the additional simulations provided in {the Supporting Information} further indicate the robustness of our proposal when the MAR assumption is violated, and also show the generalization into the cases with different types of outcomes or a relatively large number of redundant predictors in candidate models. Even for the scenarios with relatively higher missing proportions (i.e., $m=0.5$), our proposal is still applicable (results not shown). Overall, our proposed JEAIC and JEBIC outperform the other existing criteria, and JEBIC is highly recommended when the sample size is relatively large in real applications. 
	
	\begin{table}[ht]
		\centering
		\footnotesize
		\begin{minipage}{160mm}
			\caption{Performance of JEAIC and JEBIC compared with JLIC for scenarios with binary outcomes. The sample size $n=500$, $T=3$, $\rho=0.3$ across 1000 Monte Carlo datasets. Ten candidate models are considered: $\{1\}=\{x_1\}$, $\{2\}=\{x_3\}$, $\{3\}=\{x_1, x_2\}$, $\{4\}=\{x_1,x_3\}$, $\{5\}=\{x_3,x_4\}$, $\{6\}=\{x_1,x_2,x_4\}$, $\{7\}=\{x_1,x_2,x_3\}$, $\{8\}=\{x_1,x_3,x_4\}$, $\{9\}=\{x_2,x_3,x_4\}$, $\{10\}=\{x_1,x_2,x_3,x_4\}$. Note that Model \{3\}=$\{x_1,x_2\}$ with {an EXC} correlation structure is the true model. The variables $x_3$ and $x_4$ are redundant.}
			\label{table3}
			\rowcolors{2}{gray!25}{white}
			\begin{tabular}{p{0.9cm} p{0.9cm} p{0.8cm} p{0.7cm} p{0.7cm} p{0.7cm} p{0.7cm} p{0.7cm} p{0.7cm} p{0.7cm} p{0.7cm} p{0.7cm} p{0.7cm} p{0.9cm}}
				\rowcolor{gray!50}
				\toprule
				Setups &Method		&$\bfC(\bfrho)$			&	1	&	2	&	\textbf{3}	&	4	&	5	&	6	&	7	&	8	&	9	&	10	&	total	\\\midrule
				m=0.1	&	JLIC	&	AR1	&	0	&	0	&	0.03	&	0	&	0	&	0.007	&	0.007	&	0	&	0	&	0.003	&	0.047	\\
				&		&	\textbf{EXC}	&	0.006	&	0	&	\textbf{0.645}	&	0	&	0	&	0.132	&	0.147	&	0	&	0	&	0.023	&	\textbf{0.953}	\\
				&   	&	IND	&	0	&	0	&	0	&	0	&	0	&	0	&	0	&	0	&	0	&	0	&	0	\\
				
				&		&	Total	&	0.006	&	0	&	\textbf{0.675}	&	0	&	0	&	0.139	&	0.154	&	0	&	0	&	0.026	&	1	\\
				&	JEAIC	&	AR1	&	0	&	0	&	0.006	&	0	&	0	&	0.001	&	0.003	&	0	&	0	&	0.001	&	0.011	\\
				&		&	\textbf{EXC}	&	0.002	&	0	&	\textbf{0.698}	&	0	&	0	&	0.138	&	0.128	&	0	&	0	&	0.023	&\textbf{0.989}	\\
				&		&	IND	&	0	&	0	&	0	&	0	&	0	&	0	&	0	&	0	&	0	&	0	&	0	\\
				&		&	total	&	0.002	&	0	&	\textbf{0.704}	&	0	&	0	&	0.139	&	0.131	&	0	&	0	&	0.024	&	1	\\
				&	JEBIC	&	AR1	&	0	&	0	&	0.011	&	0	&	0	&	0	&	0	&	0	&	0	&	0	&	0.011	\\
				&		&	\textbf{EXC}	&	0.011	&	0	&	\textbf{0.952}	&	0	&	0	&	0.017	&	0.009	&	0	&	0	&	0	&	\textbf{0.989}	\\
				&		&	IND	&	0	&	0	&	0	&	0	&	0	&	0	&	0	&	0	&	0	&	0	&	0	\\
				&		&	total	&	0.011	&	0	&\textbf{0.963}	&	0	&	0	&	0.017	&	0.009	&	0	&	0	&	0	&	1	\\\midrule
				
				m=0.2	&	JLIC & AR1	&	0	&	0	&	0.057	&	0	&	0	&	0.011	&	0.009	&	0	&	0	&	0.001	&	0.078	\\
				&		&	\textbf{EXC}	&	0.008	&	0	&	\textbf{0.63}	&	0.002	&	0	&	0.12	&	0.137	&	0	&	0.025	&	0	&	\textbf{0.922}	\\
				&		&	IND	&	0	&	0	&	0	&	0	&	0	&	0	&	0	&	0	&	0	&	0	&	0	\\
				&		&	Total	&	0.008	&	0	&	\textbf{0.687}	&	0.002	&	0	&	0.131	&	0.146	&	0	&	0.025	&	0.001	&	1	\\
				&	JEAIC	&	AR1	&	0.001	&	0	&	0.016	&	0	&	0	&	0.002	&	0.004	&	0	&	0	&	0.001	&	0.024	\\
				&		&	\textbf{EXC}	&	0.001	&	0	&	\textbf{0.687}	&	0.001	&	0	&	0.146	&	0.12	&	0	&	0	&	0.021	&	\textbf{0.976}	\\
				&		&	IND	&	0	&	0	&	0	&	0	&	0	&	0	&	0	&	0	&	0	&	0	&	0	\\
				&		&	total	&	0.002	&	0	&	\textbf{0.703}	&	0.001	&	0	&	0.148	&	0.124	&	0	&	0	&	0.022	&	1	\\
				&	JEBIC	&	AR1	&	0.001	&	0	&	0.022	&	0	&	0	&	0	&	0	&	0	&	0	&	0	&	0.023	\\
				&		&	\textbf{EXC}	&	0.026	&	0	&	\textbf{0.922}	&	0	&	0	&	0.015	&	0.014	&	0	&	0	&	0	&	\textbf{0.977}	\\
				&		&	IND	&	0	&	0	&	0	&	0	&	0	&	0	&	0	&	0	&	0	&	0	&	0	\\
				&		&	total	&	0.027	&	0	&	\textbf{0.944}	&	0	&	0	&	0.015	&	0.014	&	0	&	0	&	0	&	1	\\
				\bottomrule
			\end{tabular}
			%		\begin{tablenotes}
			%			\footnotesize
			%			\item Ten candidate models are: $\{1\}=\{x_1\}$, $\{2\}=\{x_3\}$, $\{3\}=\{x_1, x_2\}$, $\{4\}=\{x_1,x_3\}$, $\{5\}=\{x_3,x_4\}$, $\{6\}=\{x_1,x_2,x_4\}$, $\{7\}=\{x_1,x_2,x_3\}$, $\{8\}=\{x_1,x_3,x_4\}$, $\{9\}=\{x_2,x_3,x_4\}$, $\{10\}=\{x_1,x_2,x_3,x_4\}$.    
			%		\end{tablenotes} 
		\end{minipage}
		\vspace*{-6pt}
	\end{table}

	\begin{table}[ht]
		\small
		\begin{minipage}{155mm}
			\centering
			\caption{Performance of JEAIC and JEBIC compared with JLIC for scenarios with Gaussian outcomes. The sample size $n=500$, $T=3$, $\rho=0.3$ across 1000 Monte Carlo datasets. Ten candidate models are considered: $\{1\}=\{x_1\}$, $\{2\}=\{x_2\}$, $\{3\}=\{x_1, x_2\}$, $\{4\}=\{x_1,x_3\}$, $\{5\}=\{x_1,x_3,x_{1,3}\}$, $\{6\}=\{x_1,x_2,x_{1,2}\}$, $\{7\}=\{x_1,x_2,x_3\}$, $\{8\}=\{x_2,x_3,x_{2,3}\}$, $\{9\}=\{x_1,x_2,x_3,x_{1,2},x_{1,3}\}$, $\{10\}=\{x_1,x_2,x_3,x_{1,2},x_{1,3},x_{2,3}\}$. Note that Model $\{7\}$=$\{x_1,x_2,x_3\}$ with {an EXC} correlation structure is the true model. The variable $x_4$ is redundant.}
			\label{table4}
			\rowcolors{2}{gray!25}{white}
			\begin{tabular}{p{1cm} p{1cm} p{1cm} p{0.65cm} p{0.65cm} p{0.65cm} p{0.65cm} p{0.65cm} p{0.65cm} p{0.65cm} p{0.65cm} p{0.65cm} p{0.65cm} p{0.9cm}}
				\rowcolor{gray!50}
				\toprule
				Setups &Method&$\bfC(\bfrho)$  & 1& 2& 3& 4& 5& 6& \textbf{7}& 8& 9& 10& total\\\midrule
				m=0.1& JLIC& AR1& 0& 0& 0& 0& 0& 0& 0.082& 0& 0.027& 0.012& 0.121\\
				&& \textbf{EXC}& 0& 0& 0& 0& 0& 0& \textbf{0.654}& 0& 0.141& 0.083& \textbf{0.878}\\
				&& IND& 0& 0& 0& 0& 0& 0& 0& 0& 0& 0.001& 0.001\\
				&& Total& 0& 0& 0& 0& 0& 0& \textbf{0.736}& 0& 0.168& 0.096& 1\\
				&JEAIC &AR1 &0 &0 &0 &0 &0 &0 &0.002 &0 &0 &0 &0.002 \\
				& &\textbf{EXC}  & 0& 0& 0& 0& 0& 0& \textbf{0.802}& 0& 0.124& 0.072& \textbf{0.998}\\
				& &IND &0 &0 &0 &0 &0 &0 &0 &0 &0 &0 &0 \\
				& &Total &0 &0 &0 &0 &0 &0 &\textbf{0.804}  & 0& 0.124& 0.072& 1\\
				& JEBIC& AR1& 0& 0& 0& 0& 0& 0& 0.002& 0& 0& 0& 0.002\\
				& &\textbf{EXC}  & 0& 0& 0& 0& 0& 0& \textbf{0.991}& 0& 0.005& 0.002& \textbf{0.998}\\
				& &IND &0 &0 &0 &0 &0 &0 &0 &0 &0 &0 &0 \\
				& &Total &0 &0 &0 &0 &0 &0 &\textbf{0.993}  & 0& 0.005& 0.002& 1\\
				\midrule
				m=0.2& JLIC& AR1& 0& 0& 0& 0& 0& 0& 0.163& 0& 0.034& 0.027& 0.224\\
				&& \textbf{EXC}& 0& 0& 0& 0.001& 0& 0& \textbf{0.542}& 0.002& 0.136& 0.091& \textbf{0.772}\\
				&& IND& 0& 0& 0& 0& 0& 0& 0& 0& 0.001& 0.003& 0.004\\
				&& Total& 0& 0& 0& 0.001& 0& 0& \textbf{0.705}& 0.002& 0.171& 0.121& 1\\
				&JEAIC &AR1 &0 &0 &0 &0 &0 &0 &0.01 &0 &0.006 &0.001 &0.017 \\
				& &\textbf{EXC}  & 0& 0& 0& 0& 0& 0& \textbf{0.744}& 0& 0.156& 0.083& \textbf{0.983}\\
				& &IND &0 &0 &0 &0 &0 &0 &0 &0 &0 &0 &0 \\
				& &Total &0 &0 &0 &0 &0 &0 &\textbf{0.754}  & 0& 0.162& 0.084& 1\\
				& JEBIC& AR1& 0& 0& 0& 0& 0& 0& 0.016& 0& 0.001& 0& 0.017\\
				& &\textbf{EXC}  & 0& 0& 0& 0& 0& 0& \textbf{0.975}& 0& 0.007& 0.001& \textbf{0.983}\\
				& &IND &0 &0 &0 &0 &0 &0 &0 &0 &0 &0 &0 \\
				& &Total &0 &0 &0 &0 &0 &0 &\textbf{0.991}  & 0& 0.008& 0.001& 1\\
				\bottomrule
			\end{tabular}
			%		\begin{tablenotes}
			%			\footnotesize
			%			\item {Ten candidate models are: $\{1\}=\{x_1\}$, $\{2\}=\{x_2\}$, $\{3\}=\{x_1, x_2\}$, $\{4\}=\{x_1,x_3\}$, $\{5\}=\{x_1,x_3,x_{1,3}\}$, $\{6\}=\{x_1,x_2,x_{1,2}\}$, $\{7\}=\{x_1,x_2,x_3\}$, $\{8\}=\{x_2,x_3,x_{2,3}\}$, $\{9\}=\{x_1,x_2,x_3,x_{1,2},x_{1,3}\}$, $\{10\}=\{x_1,x_2,x_3,x_{1,2},x_{1,3},x_{2,3}\}$.}
			%		\end{tablenotes}
		\end{minipage}
		\vspace*{-6pt}
	\end{table}
	
	Tables \ref{table3} and \ref{table4} summarize the comparison between our proposal and JLIC on joint selection performance when the missing probability is 0.1 or 0.2 under binary and Gaussian scenarios. All results show that JEAIC and JEBIC outperform JLIC with {higher selection rates for the true underlying model.} The improvement becomes more substantial when the outcomes are in continuous scale. In addition, with relatively larger sample size, JEBIC performs even better, {which suggests a possible advantage in controlling false positive rates}. 
	
	\section{Real Data Applications}
	\label{s:example}
	\subsection{Case 1: the Atherosclerosis Risk in Communities (ARIC) study}
	The ARIC study was designed to investigate the causes of atherosclerosis and its clinical outcomes, the trends in rates of hospitalized myocardial infarction (MI) and {coronary heart disease} (CHD) in aged 45-64 years men and women from four US communities. We select Forsyth County to identify a total of 1,036 white patients who were diagnosed with hypertension at the first examination in 1987-1989 for analysis \citep{kim2012}. {The existing literature has shown that SBP is an important risk factor for CVD risk prediction; however, the findings on its longitudinal pattern vary across studies due to several factors such as small sample size, lack of model diagnosis, limiting factors and so on \citep{muntner2015}. Here, we utilize the large epidemiological ARIC study for more exploration.} During the study period, longitudinal SBP measures were collected at approximately three-year intervals (1987-1989, 1990-1992, 1993-1995, and 1996-1998). There exist 355 dropout subjects, {leading to a monotone missing pattern}. The baseline covariates of interest are considered for exploration: age (in years), gender(1=female; 0=male), diabetes (1=fasting glucose $\geq$ 126mg/dL; 0=fasting glucose $<$ 126mg/dL), ever smoker (1=yes; 0=no), and also the examination times are coded as 1, 2, 3 and 4 for four time intervals. Before modeling, data processing is conducted, where the age variable is centered at the mean age of 54 and divided by 10 to represent a decade, and also SBP is standardized \citep{kim2012}. Also, the dropout probability $\lambda_{ij}$ is estimated from a logistic model with independent variables including all baseline covariates aforementioned and $Y_{i,j-1}$, $Y_{i,j-2}$, and $Y_{i,j-3}$.
	%\begin{equation*}
	%log(\frac{\lambda_{ij}}{1-\lambda_{ij}})=\theta_0+x_{ij1}\theta_1+x_{ij2}\theta_2+x_{ij3}\theta_3+x_{ij4}\theta_4+x_{ij5}\theta_5+Y_{i,j-1}\theta _4
	%+Y_{i,j-2}\theta _5+Y_{i,j-3}\theta _6, 
	%\end{equation*}
	
	\begin{table}[!p]	
		\footnotesize	
		\begin{minipage}{155mm}	
			\centering
			\caption{Analysis of the ARIC study based on eight candidate marginal mean regressions and three potential correlation structures. Summary results include WGEE estimates with standard errors in parentheses under {{an AR1}} ``working" correlation structure, and JEAIC, JEBIC, MLIC and QICW$_r$ for model selection. Note that for MLIC and QICW$_r$, {an EXC} correlation structure is selected based on MLICC and QICW$_r$, respectively. Notation $^\dagger$ denotes the corresponding p-value$<0.05$. }
			\label{tablesbp}
			
			\begin{threeparttable}
				\begin{tabular}{p{1.2cm} p{0.5cm} p{1.2cm}p{1.2cm}p{1.2cm}p{1.2cm}p{1.2cm}p{1.2cm}p{1.2cm}p{1.2cm}}
					\toprule
					\text{Predictors}  & $\bfC(\bfrho)$ & Model 1 & Model 2 & Model 3 & Model 4 & Model 5 & Model 6 & Model 7 & Model 8\\ \midrule
					time	&		&	0.05$^\dagger$  (0.012)	&	0.05$^\dagger$  (0.012)	&	0.05$^\dagger$  (0.012)	&	0.05$^\dagger$  (0.012)	&	0.05$^\dagger$  (0.012)	&	0.05$^\dagger$  (0.012)	&	0.05$^\dagger$  (0.012)	&	0.05$^\dagger$ (0.012)	\\
					gender	&		&	-0.15$^\dagger$ (0.054)	&	-0.10$^\dagger$  (0.050)	&		&	-0.11$^\dagger$ (0.052)	&	-0.14$^\dagger$ (0.052)	&	-0.10$^\dagger$ (0.050)	&		&	-0.13$^\dagger$ (0.052)	\\
					smoke	&		&	-0.10 (0.057)	&		&	-0.07 (0.052)	&		&	-0.11$^\dagger$ (0.055)	&		&	-0.07 (0.052)	&	-0.11$^\dagger$ (0.055)	\\
					age	&		&		&	0.36$^\dagger$  (0.045)	&	0.37$^\dagger$  (0.045)	&		&	0.37$^\dagger$  (0.045)	&	0.36$^\dagger$  (0.045)	&	0.36$^\dagger$  (0.046)	&	0.36$^\dagger$  (0.045)	\\
					diabetes	&		&		&		&		&	0.14 (0.076)	&		&	0.09 (0.078)	&	0.10 (0.076)	&	0.09 (0.077)	\\
					\midrule
					
					JEAIC	&	AR1	&	123.46	&\textbf{55.24}	&	59.58	&	117.78	&	56.26	&	58.04	&	62.51	&	58.95	\\
					&	EXC	&	129.37	&	71.02	&	73.19	&	121.8	&	70.37	&	70.51	&	72.66	&	69.68	\\
					&	IND	&	922.75	&	789.34	&	781.46	&	896.88	&	798.1	&	780.41	&	785.85	&	791.26	\\
					\midrule
					
					JEBIC	&	AR1	&	148.18	&	\textbf{79.96}	&	84.29	&	142.5	&	85.92	&	87.7	&	92.17	&	93.55	\\
					&	EXC	&	154.09	&	95.73	&	97.91	&	146.52	&	100.03	&	100.17	&	102.32	&	104.28	\\
					&	IND	&	942.52	&	809.11	&	801.23	&	916.66	&	822.81	&	805.13	&	810.57	&	820.91	\\
					\midrule
					MLIC	&	EXC	&	5118	&	\textbf{5037.9}	&	5040.7	&	5117.2	&	5040.1	&	5042.5	&	5044.9	&	5044.8	\\
					\midrule
					
					QICW$_r$	&	EXC	&	5114.7	&	\textbf{5035.1}	&	5037.9	&	5113.5	&	5036.6	&	5038.6	&	5041	&	5040.1	\\
					\bottomrule
					
				\end{tabular}
				
				%			\begin{tablenotes}
				%				
				%				\footnotesize
				%				
				%				\item $^\dagger$ denotes the corresponding p-value$<0.05$.
				%				\item $^{*}$: the EXC correlation structure is selected based on MLICC or QICW$_r$. 
				%				
				%			\end{tablenotes}
				
			\end{threeparttable}
			
		\end{minipage}
		
	\end{table}
	
	Table \ref{tablesbp} summarizes the results with {the boldface values} indicating that the information criterion is the smallest among possible candidate models. From Table \ref{tablesbp}, {Model 2 with {an AR1} correlation structure is selected by JEAIC, JEBIC, while Model 2 with {an EXC} correlation structure is selected by MLIC/MLICC and QICW$_r$.} Thus, marginal mean regression is selected consistently; however, the discrepancy in the selected correlation structures based on different criteria shows the necessity and importance to utilize more robust and reliable information criteria. Furthermore, we check the empirical pairwise correlations between times, and a decreasing trend is shown when time gap becomes larger, indicating our selection is reasonable and valid. The final selected model, Model 2, includes three variables: time, gender, and age, which all have significant effects on SBP.
	%\vspace{0.1in}
	\subsection{Case 2: the National Institute of the Mental Health Schizophrenia (IMPS) Study}
	To {further evaluate} our proposal for categorical outcomes, we consider the data from the IMPS study {that includes 293 patients in the treatment group who were given drugs chlorprom azine, fluphenazine, or thioridazine as treatment} and 93 patients in placebo group \citep{gibbons1994}. For each patient, the severity of schizophrenia disorder (IMPS79) was measured (range: 0-7) at week 0, 1, 3, 6 (time=$\sqrt{\text{week}}$). Here, we define $Y = 1$ if IMPS $\geq 4$; otherwise, $Y = 0$. The goal is to investigate treatment effect (drug=1 for treatment; 0 for placebo) and sex (1=male; 0=female) on $Y$. The dropout probability $\lambda_{ij}$ is estimated from a logistic regression with the predictors $drug_{ij}$, $sex_{ij}$, $time_{ij}$, $Y_{i,j-1}$, $Y_{i,j-2}$, and $Y_{i,j-3}$. 
	%\begin{equation*}
	%	log(\frac{\lambda_{ij}}{1-\lambda_{ij}})=\theta_0+Drug_{ij}\theta_1+Time_{ij}\theta_2+Sex_{ij}\theta_3+Y_{i,j-1}\theta _4
	%	+Y_{i,j-2}\theta _5+Y_{i,j-3}\theta _6, 
	%\end{equation*}
	\begin{table}[!p]	
		\footnotesize	
		\begin{minipage}{155mm}	
			\centering
			\caption{Analysis of the IMPS study based on six candidate marginal mean regressions and three correlation structures. Summary results include WGEE estimates with standard errors in parentheses under {an AR1} ``working" correlation structure, and JEAIC, JEBIC, MLIC and QICW$_r$ for model selection. Note that for MLIC and QICW$_r$, {an AR1} correlation structure is selected based on MLICC and QICW$_r$, respectively. Notation $^\dagger$ denotes the corresponding p-value$<0.05$. }
			\label{tableexa}
			
			\begin{threeparttable}
				\begin{tabular}{p{1.5cm} p{0.5cm} p{1.6cm} p{1.6cm} p{1.6cm} p{1.6cm} p{1.6cm} p{1.6cm} }
					\toprule
					\text{Predictors}  & $\bfC(\bfrho)$ & Model 1 & Model 2 & Model 3 & Model 4 & Model 5 & Model 6 \\ \midrule
					
					time & & -1.339 (0.081)$^\dagger$ &  & -1.372 (0.084)$^\dagger$ & -1.166 (0.208)$^\dagger$ & -1.372 (0.084)$^\dagger$ & -1.180 (0.239)$^\dagger$ \\ 
					
					drug & &  &  -0.618 (0.182)$^\dagger$ & -0.854 (0.236)$^\dagger$ & -0.357 (0.438)  & -0.860 (0.237)$^\dagger$ & -0.524 (0.492)    \\
					
					sex & & &  &  &  & 0.116 (0.184) & -0.188 (0.494)    \\
					
					time*drug & &  &  &  & -0.256 (0.271) &  & -0.252 (0.229)    \\
					
					time*sex & & &  &  &  &  & 0.023 (0.171)    \\
					
					sex*drug & & &  &  &  &  & 0.345 (0.460)     \\\midrule
					
					JEAIC & AR1 & 27.55 & 398.50 & \textbf{16.08} & 17.55 & 17.64 & 22.63 \\
					
					& EXC & 94.52 & 491.44 & 90.70 & 91.87 & 94.14 & 101.78 \\
					
					& IND & 223.56 & 496.46 & 209.77 & 210.76 & 209.69 & 212.86 \\ \midrule
					
					JEBIC & AR1 & 39.42 & 410.37 & \textbf{31.91} & 37.33 & 37.42 & 54.28 \\
					
					& EXC & 106.38 & 503.31 & 106.53 & 111.65 & 113.92 & 133.43 \\
					
					& IND & 231.48 & 504.37 & 221.64 & 226.59 & 225.51 & 240.56 \\\midrule
					
					MLIC  & AR1 & 261.9 & 321.5 & \textbf{255.8} & 256 & 256.5 & 257.5 \\ \midrule
					
					QICW$_r$& AR1 & 1554.8 & 1872.2 & 1529.6 & \textbf{1529.5} & 1532.7 & 1537.1 \\
					
					\bottomrule
					
				\end{tabular}
				
				%			\begin{tablenotes}
				%				
				%				\footnotesize
				%				
				%				\item $^\dagger$ denotes the corresponding p-value$<0.05$.
				%				\item $^{*}$: the AR1 correlation structure is selected based on MLICC or QICW$_r$. 
				%			\end{tablenotes}
				
			\end{threeparttable}
			
		\end{minipage}
		
	\end{table}
	Table \ref{tableexa} summarizes the results of model fitting and comparisons. Note that previous work {has shown that {an AR1} correlation structure} is preferred based on MLICC; thus MLIC and QICW$_r$ are calculated given this AR1 selection. %however, we still consider all combinations for JEAIC and JEBIC for joint selection on marginal mean and correlation structures. 
	Table \ref{tableexa} shows that Model 3 is selected as the best candidate model based on JEAIC, JEBIC, and MLIC because of the minimum values among all six candidate models. However, QICW$_r$ selects Model 4 as the best one even though the value is slightly lower than that {of Model 3}. Lastly, the final selected model, Model 3, includes two variables, time and drug, which both have significant effects on the risk of severe schizophrenia disorder.
	
	\section{Discussion}
	\label{s:discussion}
	In this paper, we heuristically introduce two innovative information criteria, JEAIC and JEBIC, for longitudinal data with dropout missingness under MAR. {The proposed criteria are evaluated in both theoretical and numerical studies with better performance compared to MLIC, QICW$_r$ and JLIC under a variety of scenarios}. In particular, the expected quadratic loss distance based upon which MLIC and JLIC are derived is {a model-free criterion}, which only measures how well the estimated means approximate to the population means but without identifying {the true mean structure \citep{ye1998}.} {Thus, it might not be easy to distinguish two mean structures, which are both close to the true mean under finite samples}. On the other hand, QICW$_r$ modifies QIC and implements correlation structure selection based on so-called ``more informative" penalty term \citep{gosho2016}. {However, it is unclear in theory whether and how correctly} specifying a ``working" correlation structure will intrinsically minimize the penalty term in QICW$_r$. {In contrast, our proposed JEAIC and JEBIC} are based on empirical likelihood, which are distribution-free and efficiently driven by observed data and informative estimating equations. This accordingly provides scientific sense why our empirical-likelihood-based criteria would have outperformance, assuming that the true underlying model is nested within the full estimating equations. {Our approach is easy to be implemented in software with the code available in the Supporting Information}. Also, extensive simulations show that our proposed criteria perform computationally efficient and are flexible to be extended for more complicated scenarios, indicating the potential for wide application. 
	
	Despite the aforementioned advantages brought up from JEAIC and JEBIC, {there is still substantial work} for further evaluation or improvement, for instance, selection stability to account for sampling variability may need more check via extensive simulation studies using a bootstrap approach. Also, two other potential extensions may include: 1) to accommodate more general missing patterns such as intermittent missingness; 2) to consider the missingness on some time-dependent covariates or high-dimensional predictors (i.e., gene expression data) \citep{chen2010}, which is also commonly encountered in practice nowadays. Therefore, how to generalize our proposal and accurately perform joint model selection under these scenarios {still needs to be explored}.
	
	%\section{Software}
	%\label{sec5}
	
	%Software in the form of R code, together with a sample
	%input data set and complete documentation is available on
	%request from the corresponding author (eaheron@tcd.ie).
	
	\vspace*{-8pt}
	\backmatter
	\section*{Acknowledgments}
	\label{ack}
	Wang's research was partially supported by Grant UL1 TR002014 and KL2 TR002015 from the National Center for Advancing Transnational Sciences (NCATS). The content is solely the responsibility of the authors and does not represent the official views of the National Institute of Health, the National Science Foundation and other research sponsors. 
	
	\section*{Supporting information}
	{The Web Appendices of proofs, additional tables for the simulation studies, the IMPS data example analyzed in Section 4.2 and R codes implementing our method are available with this article at the Biometrics website on Wiley Online Library.}
	
\bibliographystyle{biom} 
\bibliography{biomsample_bib}

\label{lastpage}

\end{document}